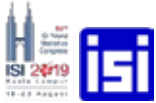

# Flipping A Statistics Classroom for Pre-Service English Language Teachers

Yosep Dwi Kristanto[1*]; Russasmita Sri Padmi[2]

[1] Universitas Sanata Dharma, Yogyakarta, Indonesia
[2] SEAMEO QITEP in Mathematics, Yogyakarta, Indonesia

*Corresponding author: yosepdwikristanto@usd.ac.id

**Abstract:**
Flipped classroom approach has gained attention for educational practitioners and researchers in recent years. In contrast with traditional classroom, in flipped classroom, students gather basic knowledge out of class, so that class time can be used by them to engage in active learning experience. While rich body of research has been investigated the effectiveness of flipped classroom, yet little is known about student's perspective towards the implementation of flipped classroom, especially for non-mathematics students that enrol statistics course. This study aims to describe pre-service English language teacher's perceptions with regard to the implementation of the learning approach in Statistics for Research course. Online questionnaire is employed in capturing student's perceptions. Students were generally positive regarding the implementation of flipped classroom. Although, the students still had difficulty with regard to online learning environment, they felt that many features of flipped classroom are helpful for their learning experiences.

## 1. Introduction

Recently flipped classroom pedagogical approach acquire popularity in educational research and practice. This learning approach flip the events that usually occurs in classroom to be happen out of class, and vice versa (Lage, Platt, & Treglia, 2000). Educational practitioner and researchers implemented the approach in various models (e. g., Lai & Hwang, 2016; Lee, Lim, & Kim, 2017). However, the basic idea of flipped classroom is to move basic knowledge acquisition out of class, so that students have preparation to deepen their knowledge and understanding when they are in class. In flipped classroom, students typically watch videos and take quiz in their own learning environments (e.g. home, library, boarding house, etc.) and then, in class, they will engage in group activities or practicum.

Rich body of research has investigated the effectiveness of flipped classroom. In nursing education, Betihavas, Bridgman, Kornhaber, and Cross (2016) found that flipped classroom has impact on improving academic performance outcome. Some literature reviews (e. g., Long, Logan, & Waugh, 2016); O'Flaherty & Phillips, 2015) also give the evidences of its effectiveness. However, little is known about student's perceptions towards flipped classroom, particularly for non-mathematics students that is "forced" to enrol mathematics course. Therefore, the aims of the present study are to describe pre-service English language teacher's perceptions with regard to the implementation of the learning approach in Statistics for Research course.

The manuscript is organized as follows: the implementation of flipped classroom is described in Section 2; methodology of the current study is described in Section 3; statistical summaries of the data and their corresponding analysis are presented in Section 4; and discussion of the results and concluding remark appear in Section 5.

## 2. The description of the study

In Statistics for Research course, there was an introductory lecture in week 1, to explain the course syllabus, then followed by two subsequent terms that conducted in different learning approach. Flipped classroom approach was implemented in first term of the course, spanning 4 weeks, which was taught by the first author. In the first until fourth weeks of the first term, students learned basic descriptive statistics, measures of central tendency, measures of dispersion, and normal curve respectively. The course, then, was delivered using traditional approach in the second term in which students learned inferential statistics.

*Pre-class activity.* In general, flipped classroom approach has two learning stages, namely pre-class activity and in-class activity. In pre-class activity, students listened podcasts, watched videos, took online quizzes, and did self-assessment through the Exelsa (http://exelsa2012.usd.ac.id/), Moodle-based learning management system (LMS) provided by the university. The podcasts provided introduction regarding the topic to be learned. The videos were used to introduce concepts, demonstrate procedures, and illustrate with real life scenarios (Lim, & Wilson, 2018). The online quizzes were aimed to facilitate students in recalling their knowledge right after they watched the videos. The self-assessment was aimed to facilitate students' reflection about their learning. Those learning materials were arranged so that they were sequential.

Table 1.
*Number of audio plays in Soundcloud*

| Audio | Audio title | Duration in minutes | Soundcloud plays |
|---|---|---|---|
| 1 | Basic Descriptive Statistics: An Introduction | 0:57 | 21 |
| 2 | Measures of Central Tendency: An Introduction | 1:41 | 11 |
| 3 | Measures of Dispersion: An Introduction | 2:31 | 9 |
| 4 | The Normal Curve: An Introduction | 1:11 | 7 |

The podcasts were produced using smartphone's audio recording feature. This production technique usually obtained low quality audios. Therefore, the recorded audios then were edited using an audio editor software in order to produced more clear sounds. After the audios were settled up, they were uploaded in Soundcloud. Details of the podcasts used in the present study were shown on Table 1.

Table 2.
*The statistics summaries of videos in each topic*

| Topic | Number of videos | Duration average (in minutes) | YouTube views average | YouTube time watched average (in hours) |
|---|---|---|---|---|
| Basic Descriptive Statistics | 8 | 6.65 | 36.13 | 1.18 |
| Measures of Central Tendency | 5 | 6.74 | 25.8 | 1.09 |
| Measures of Dispersion | 4 | 6.46 | 45 | 1.7 |
| The Normal Curve | 3 | 6.71 | 47 | 2.11 |

Screencast was employed in producing the videos. Screencast is a digital record of computer screen output that often contains audio narration (Udell, 2005). The use of screencast has several benefits for students' learning. Students can view the screencast at their own convenience. They also will have more rich learning experiences when watching the combination of images and sounds compared with tradition textbooks (Sugar, Brown, & Luterbach, 2010). Both of video and audio editor software were used to produce the screencasts. The final videos then were uploaded to YouTube and embedded in LMS. The accompanying PowerPoint slides for the videos also uploaded to LMS so that students can download and use them to learn the covered topics. In total, 20 videos were provided to students that covered 4 main topics. Details of the videos is shown in Table 2 and an example of the video is shown in Figure 1.

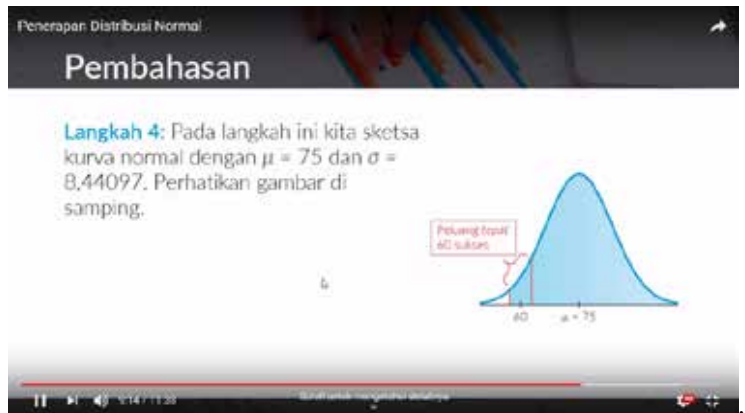
Figure 1. Screencast in the topic of Application of Normal Distributions

*In-class activity.* Aside from pre-class activity, in-class activity is another critical component of flipped classroom approach. In-class activity consists interactive group activities that addresses student-centred learning (Bishop & Verleger, 2013). In 3 out of 4 face-to-face meetings, the instructor began the class with peer instruction for approximately 15 minutes. This peer instruction aimed to help students recalling their prior knowledge and understanding about the materials they learned in pre-class activity (Crouch & Mazur, 2001). In this phase, the instructor used Desmos in presenting a multiple-choice question in front of class and poll the students' answer. When there was disagreement in students' answer, the instructor urged them to convinced their fellow students in order to agree with their answer by explaining the underlying reasoning. Finally, the instructor continued the learning activity by conducting poll again with the same question. If necessary, the instructor gave a brief explanation about the answer of the question. One meeting did not employed the peer instruction in the beginning. Rather, the instructor directly gave a brief lecture about the topic to be learned.

In the main body of lesson in face-to-face sessions, the present study uses problem-based learning as in-class activity for 4 weeks implementation of flipped approach. In each face-to-face meeting students were given a set of problems to be solved in group. When students work in group, the instructor walked around the class to check whether the students had questions. The instructor also gave questions to students in order to know their understanding and to challenge the students' thinking. This instructor's role were important principles in implementing problem-based learning (Savery & Duffy, 1995). In the end of the meeting, students should submit their works through the LMS.

### 3. Methodology

The present study takes descriptive methods in exploring students' learning experiences in flipped classroom by using their perceptions. The study was conducted in a Statistics for Research course in the department of English language education at a private university in Yogyakarta. Thirty third-year students were enrolled in the course, of which, 5 were males and 25 were females. After the implementation of flipped approach, students were asked to take online survey that can be accessed through the LMS. The online survey consisted 14 close questions and 2 open questions. The close questions were 5-point scale that asked the agreement of the given statement, whereas the open questions asked students' opinion regarding their favourite and unpleasant experiences in flipped classroom.

The quantitative data of student's perceptions were analyse using the Statistical Package for Social Sciences (SPSS) version 24. Statistical summaries of student's respond in each questionnaire item were calculated. In analysing student's written respond regarding their positive and negative learning experiences in flipped classroom, Atlas.ti version 7 was employed. The student's written responds were categorized with codes and then the occurrence of the categories were calculated to see how grounded the categories is.

### 4. Result

Table 3 provides the means, standard deviations, standard errors, and 95% confidence intervals for the mean of student perceptions towards flipped classroom.

Table 3.
*Student's perceptions towards flipped classroom*

| Statement | M | SD | SE Mean | 95% CI |
|---|---|---|---|---|
| Pre-class perceptions | | | | |
|     The videos were useful to my learning. | 4.034 | 0.778 | 0.145 | (3.738, 4.331) |
|     The quizzes in online lesson were useful to my learning. | 3.793 | 0.62 | 0.115 | (3.557, 4.029) |
| In-class perceptions | | | | |
|     The lesson review in the beginning of every face-to-face meeting was useful to my learning. | 4.069 | 0.704 | 0.131 | (3.801, 4.337) |
|     The due date and time for in-class assignments were set reasonably | 3.138 | 1.06 | 0.197 | (2.735, 3.541) |
|     The in-class group questions were thought-provoking and helped me to deepen my knowledge. | 3.793 | 0.774 | 0.144 | (3.499, 4.087) |
|     My in-class discussions with peers and the instructor helped me learn. | 4.207 | 0.774 | 0.144 | (3.913, 4.501) |
|     The class time is structured effectively for my learning. | 3.69 | 1.004 | 0.186 | (3.308, 4.071) |
|     The class time is critical to my learning. | 3.69 | 0.85 | 0.158 | (3.367, 4.013) |
|     The structure of this flipped class supports my learning in and out of class. | 3.276 | 0.96 | 0.178 | (2.911, 3.641) |
|     Having to communicate mathematics in class helped me learn the concepts better. | 3.69 | 0.891 | 0.165 | (3.351, 4.028) |

Student's written opinions with regard to their positive and negative experiences in flipped classroom are classified into several categories. Table 4 shows the number of occurrences of the category in student's positive or negative experiences and the examples of student's written opinion related with their experience.

Table 4.
*The number of occurrences of student's opinion in their learning experience*

| Student's opinion category | Positive Experience | Negative Experience | Example of student's opinion |
|---|---|---|---|
| Statistical Topic | 9 | 7 | I enjoy to study this course because it can help me to answer research question or in skripsi/thesis. The materials so interesting and make me fun to study it. |
| Group Discussion | 9 | 1 | My favourite experience is when we learn and discuses together with my friends. It makes me interested to follow this lesson even though it is very difficult. |
| Internet Connection | 0 | 7 | I had many troubles when assessing online class especially internet connection. |
| Instructor Role | 5 | 0 | … the teacher gives us example and make students answer the question. |
| Online Environment | 0 | 4 | I prefer to use a traditional method that classroom activity should give a learning and also practice in a same time. |
| Peer Instruction | 4 | 0 | I like the interactive learning platform: Desmos. |
| Time Management | 0 | 3 | I need more time to understand the material, and i cannot think clearly when I need to submit the exercise in short time. |
| Class Structure | 0 | 1 | When I forgot to watch the video before class because I always come home late after I sing at the |
|  |  |  | night before. |
| Language | 0 | 1 | The lecture didn't use English. |
| Result-Based | 0 | 1 | The class doesn't encourage me to do the assignment or quiz, etc. Because it is more result based, not process based. All we know just we submit our answer, but I actually don't have the passion to do it. |
| Autonomy | 1 | 0 | I can learn a review the material by myself. |
| Statistical Technology | 1 | 0 | I can learn how to use Excel, so it helps me to count and analyse the data quickly. |
| Video | 1 | 0 | Watching the videos at home is more enjoyable because it stimulates my curiosity |

## 5. Discussion and Conclusion

The aim of the current study was to explore student's perceptions toward flipped classroom. The result suggests that the students, in general, have positive perceptions towards flipped classroom approach. They perceived pre-class activity as useful phase in their learning experiences. Specifically, they felt that the use of videos can enhance their learning in their own learning environments. Long, Logan, and Waugh (2016) have similar findings regarding the use of video in flipped classroom.

The students also have positive attitudes toward in-class activity. They perceived that peer instruction in the beginning face-to-face meeting motivated them in learning statistics. This is also the case for group discussion. Instructor role in guiding and tutoring students has positive impact to students' learning as well. These findings are the evidence of what Sams and Bergmann (2012) suggest about flipped classroom. Flipped classroom optimize the use of time for learning in class time. Since students had learned basic knowledge about domain specific contents in pre-class activity, the instructor has more time to guide students in learning more advanced and engaging materials.

However, the students also have negative perceptions as well when the flipped classroom was not accompanied with adequate infrastructure, such as internet connection. The internet connection is critical requirement for students to access all materials. Time management also has important role in conducting flipped classroom. It was the case of the current study. The students regarded that they did not have enough time in solving all problems to be presented in face-to-face meeting. This finding in line with Prober and Khan (2013) findings, but in contrast with Lai and Hwang's (2016) self-regulated flipped classroom results. In their self-regulated flipped classroom, student uses self-regulated monitoring system track their learning strategies.

The findings of the current study, albeit anecdotal, suggest that flipped classroom has the promise in orchestrating active learning environments without losing time in covering basic learning materials. Engaging students with multimedia in pre-class activity provide learning environment in which they can learn essential course material to prepare them for in-class time. Therefore, the instructor can focus to deepen students' knowledge and understanding when the students come to class with active and engaging activities. This learning approach may serve a significant promise for successful implementation in Statistics course for non-mathematics students who consider Statistics as, referring Cobb's (2007) phrase, "tyranny of the computable."

Admittedly, much more research is needed in supporting the current study with regard to the effectiveness of flipped classroom from student's perspective. The current study shows student's perceptions toward flipped classroom approach, but more studies are still needed in investigating the approach's impact on student's learning.

**References**
1. Bergmann, J., & Sams, A. (2012). *Flip your classroom: Reach every student in every class every day*. International society for technology in education.
2. Betihavas, V., Bridgman, H., Kornhaber, R., & Cross, M. (2016). The evidence for 'flipping out': a systematic review of the flipped classroom in nursing education. *Nurse Education Today*, **38**, 15-21. https://doi.org/10.1016/j.nedt.2015.12.010
3. Bishop, J. L., & Verleger, M. A. (2013). The flipped classroom: A survey of the research. In *ASEE national conference proceedings, Atlanta, GA* (Vol. 30, No. 9, pp. 1-18).

4. Cobb, G. W. (2007). The Introductory Statistics Course: A Ptolemaic Curriculum? *Technology Innovations in Statistics Education*, **1**(1). Retrieved from https://escholarship.org/uc/item/6hb3k0nz
5. Crouch, C. H., & Mazur, E. (2001). Peer instruction: Ten years of experience and results. *American Journal of Physics*, **69**(9), 970-977. https://doi.org/10.1119/1.1374249
6. Lage, M. J., Platt, G. J., & Treglia, M. (2000). Inverting the classroom: A gateway to creating an inclusive learning environment. *The Journal of Economic Education*, **31**(1), 30-43. https://doi.org/10.2307/1183338
7. Lai, C. L., & Hwang, G. J. (2016). A self-regulated flipped classroom approach to improving students' learning performance in a mathematics course. *Computers & Education*, **100**, 126-140. https://doi.org/10.1016/j.compedu.2016.05.006
8. Lee, J., Lim, C., & Kim, H. (2017). Development of an instructional design model for flipped learning in higher education. *Educational Technology Research and Development*, **65**(2), 427-453. https://doi.org/10.1007/s11423-016-9502-1
9. Lim, K. H., & Wilson, A. D. (2018). Flipped Learning: Embedding Questions in Videos. *Mathematics Teaching in the Middle School*, **23**(7), 378-385. https://doi.org/10.5951/mathteacmiddscho.23.7.0378
10. Lo, C. K., & Hew, K. F. (2017). A critical review of flipped classroom challenges in K-12 education: possible solutions and recommendations for future research. *Research and Practice in Technology Enhanced Learning*, **12**(1), 4. https://doi.org/10.1186/s41039-016-0044-2
11. Long, T., Logan, J., & Waugh, M. (2016). Students' perceptions of the value of using videos as a pre-class learning experience in the flipped classroom. *TechTrends*, **60**(3), 245-252. https://doi.org/10.1007/s11528-016-0045-4
12. O'Flaherty, J., & Phillips, C. (2015). The use of flipped classrooms in higher education: A scoping review. *The Internet and Higher Education*, **25**, 85-95. https://doi.org/10.1016/j.iheduc.2015.02.002
13. Prober, C. G., & Khan, S. (2013). Medical education reimagined: a call to action. *Academic Medicine*, *88*(10), 1407-1410. https://doi.org/10.1097/acm.0b013e3182a368bd
14. Savery, J. R., & Duffy, T. M. (1995). Problem based learning: An instructional model and its constructivist framework. *Educational Technology*, **35**(5), 31-38.
15. Sugar, W., Brown, A., & Luterbach, K. (2010). Examining the anatomy of a screencast: Uncovering common elements and instructional strategies. *The International Review of Research in Open and Distributed Learning*, **11**(3), 1-20. https://doi.org/10.19173/irrodl.v11i3.851
16. Udell, J. (2005). What is screencasting? Retrieved from http://digitalmedia.oreilly.com/pub/a/oreilly/digitalmedia/2005/11/16/what-isscreencasting.html?page=2#heading2